\documentclass[%
 reprint,superscriptaddress,
 twocolumns,
nofootinbib,
 amsmath,amssymb,
 aps,
 prl,
floatfix,
]{revtex4-1}
\usepackage{graphicx}

\begin{document}


\title{Shortcuts to equilibrium with a levitated particle in the underdamped regime}
\author{Damien Raynal}
\author{Timoth\'{e}e de Guillebon}%
\affiliation{Universit\'{e} Paris-Saclay, ENS Paris-Saclay, CNRS, CentraleSupélec, LuMIn, 91405 Orsay Cedex, France}
\author{David Gu\'{e}ry-Odelin}
\affiliation{Universit\'{e} Paul Sabatier - Toulouse 3, CNRS, LCAR, 31062 Toulouse Cedex 9, France}
\author{Emmanuel Trizac}
\affiliation{Universit\'{e} Paris-Saclay, CNRS, LPTMS, 91405 Orsay Cedex, France}
\affiliation{Univ. Lyon, ENS de Lyon, F-69342 Lyon, France}
\author{Jean-S\'{e}bastien Lauret}%
\author{Lo\"{i}c Rondin}%
\email{loic.rondin@universite-paris-saclay.fr}
\affiliation{Universit\'{e} Paris-Saclay, ENS Paris-Saclay, CNRS, CentraleSupélec, LuMIn, 91405 Orsay Cedex, France}

\date{\today}

\begin{abstract}
We report on speeding up equilibrium recovery in the previously unexplored general case of the underdamped regime using an optically levitated particle. We accelerate the convergence towards equilibrium by an order of magnitude compared to the natural relaxation time. We then discuss the efficiency of the studied protocols, especially for a multidimensional system. These results pave the way for optimizing realistic nanomachines with application to sensing and developing efficient nano-heat engines. 
\end{abstract}

\maketitle

In the quest to achieve better control over physical systems in the classical, quantum and stochastic realms, the ability to perform transformations between equilibrium states is paramount. The opportunity to realize such transformations at high speed could then improve the system's efficiency. 
For instance, in the context of stochastic thermodynamics, shortcuts to equilibrium protocols have been proposed to optimize force sensors~\cite{LeCunuder2016}, nano-heat engines~\cite{Albay2020APL} or computing~\cite{Boyd2022JSP}. These applications have triggered significant works devoted to one-dimensional shortcut protocols, where the relaxation path of the system is designed to reach equilibrium in an arbitrarily short time $t_f$, faster than the natural relaxation time $t_\text{relax}$ that governs the time evolution after a sudden change of the system parameters~\cite{Guery-Odelin2019RMP, Martinez2016NP,Li2017PRE, Chupeau2018PRE, Dago2020SP}. For instance, protocols accelerating isothermal compression and expansion have been experimentally demonstrated for overdamped systems~\cite{Martinez2016NP,PhysRevResearch.2.012012,Albay2020APL}. However, to understand the fundamental nature of the shortcuts to equilibrium protocols, one must address the more general underdamped regime, where the inertia of the system cannot be neglected and for which the position, the velocity, and their correlation must be taken into account~\cite{Li2017PRE,Chupeau_2018}. 
The underdamped regime gives access to non-thermal states, that are ideally suited for studying nonequilibrium fluctuations for transitions between arbitrary steady states~\cite{Gieseler14}. It also provides equilibrium information from nonequilibrium measurements~\cite{PhysRevE.103.032146}, and is essential for temperature-changing transitions~\cite{Martinez2015,Jun2021PRR}. 
Beyond these fundamental questions, extending shortcuts to equilibrium to the underdamped regime is also essential to improve nano-mechanical systems that usually operate in this regime, or to take advantage of the control that one can enforce on weakly damped systems,  and which may lead for example to the development of all-optical nano-heat-engines~\cite{Dechant2015PRL}. 
A first step toward answering this question has been taken with the demonstration of accelerated transport protocols using an underdamped micro-mechanical oscillator~\cite{LeCunuder2016}. Nevertheless, in this one-dimensional example, only the average system position is engineered, and the resulting protocol is oblivious to fluctuations. 
More subtle protocols are required in general to control the system's position and velocity standard deviations simultaneously, including more spatial dimensions. This is a much harder task, which should take into account fluctuations and the coupling between degrees of freedom~\cite{Chupeau_2018}. 
In that context, extending experimentally underdamped shortcut to equilibrium to more general protocols can benefit from the recent developments on optically levitated particles~\cite{Gonzalez-Ballestero2021S,Ciampini2021ACPP,Gieseler2018E}, which provide excellent tools to track particle dynamics and where the system damping can be easily tuned~\cite{Rondin2017NN}.

In this work, we implement the first experimental acceleration of a harmonic expansion in the underdamped regime. We also demonstrate that accurate measurement of levitated nanoparticles' position and velocity makes it perfectly fitted to monitor the out-of-equilibrium dynamics of stochastic systems, by allowing to easily follow dynamics in the entire phase space, which is not the case in the overdamped regime. Finally, we discuss the robustness of shortcuts to equilibrium protocols by studying the particle relaxation for its different degrees of freedom. 

%
\begin{figure}[!htb]
\includegraphics[scale=.5]{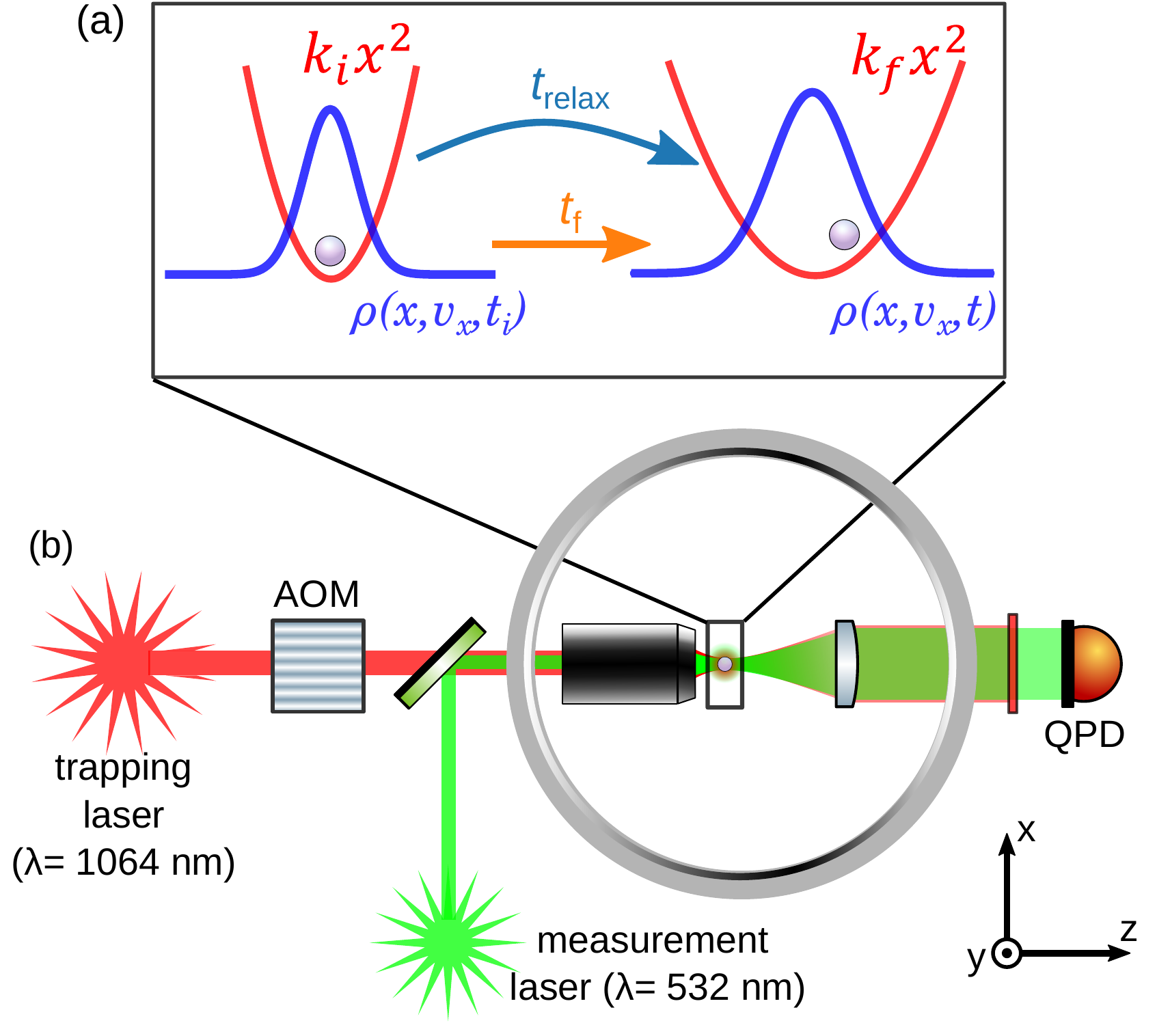}
\caption{(a) Shortcut to equilibration for a trap expansion. The trap stiffness is changed from $k_i$ to $k_f$. Using shortcut protocols, equilibrium is reached in a target time $t_f$ shorter than the natural relaxation time of the system $t_\text{relax}$. The equilibration dynamics can be tracked through the probability distribution function $\rho (x,v_x,t)$. (b) A NIR laser beam traps a silica nanoparticle at the focus of a high-NA microscope objective. The NIR laser power is controlled with an AOM (acousto-optic modulator). The particle dynamics is measured with a green laser and a quadrant photodiode.}
\label{fig:setup}
\end{figure}
Figure \ref{fig:setup} shows a sketch of the experimental setup. A 73~nm silica particle is trapped by an optical tweezer made from a high-power near-infrared (NIR) laser beam. The particle dynamics is measured with a common path interferometer using an ancillary green laser beam and a quadrant photodetector~\cite{Gieseler2012}. 
The potential experienced by the particle is well approximated by a 3D harmonic potential, whose transverse trap stiffness, $k^x_\text{trap}$, can be modified by changing the light intensity of the tweezer. We study an expansion of the potential corresponding to a change of the trap stiffness from $k_i$ to $k_f<k_i$. The objective is to develop a protocol acting on a control parameter of the system, here the trap stiffness, to accelerate the equilibration of the system in a chosen time $t_f$, shorter than the system's natural relaxation time $t_\text{relax}$.
The three eigen angular frequencies of the harmonic trapping are non-degenerate. We measure $\omega_{x}/2\pi\approx$300~kHz, $\omega_{y}/2\pi\approx$250~kHz, and $\omega_{z}/2\pi\approx$90~kHz.  The trap stiffness along each of these axes $q=\{x,y,z\}$ is $k^q_\text{trap}=m\omega_q^2$, where $m$ is the particle mass; it is then directly proportional to the NIR trapping laser power $P_\text{las}$~\cite{Gieseler2012}. 
Using an acousto-optic modulator, we can dynamically tune $P_\text{las}$ and thus finely control the trap stiffness.
The trapping setup is enclosed inside a vacuum chamber to control the particle's interaction with its environment. Indeed, the system damping rate $\Gamma$ is directly proportional to the gas pressure $p_\text{gas}$ inside the chamber~\cite{Rondin2017NN}. Here, we are specifically interested in the underdamped regime. Thus, we set a gas pressure of $p_\text{gas}=5$~hPa, corresponding to a reduced damping rate $\Gamma/2\pi \approx3$~kHz, such that the underdamped condition $\Gamma < \omega_{q}$ is fulfilled for the three axes  $q=\{x,y,z\}$~\cite{SI}. Nevertheless, note that our experimental setup allows to readily extend the presented work to a wide range of damping rates, and notably to the overdamped regime~\cite{SI}. 
More details about the experimental setup can be found in~\cite{SI}.

\begin{figure*}[!htbp]
\includegraphics[scale=.4]{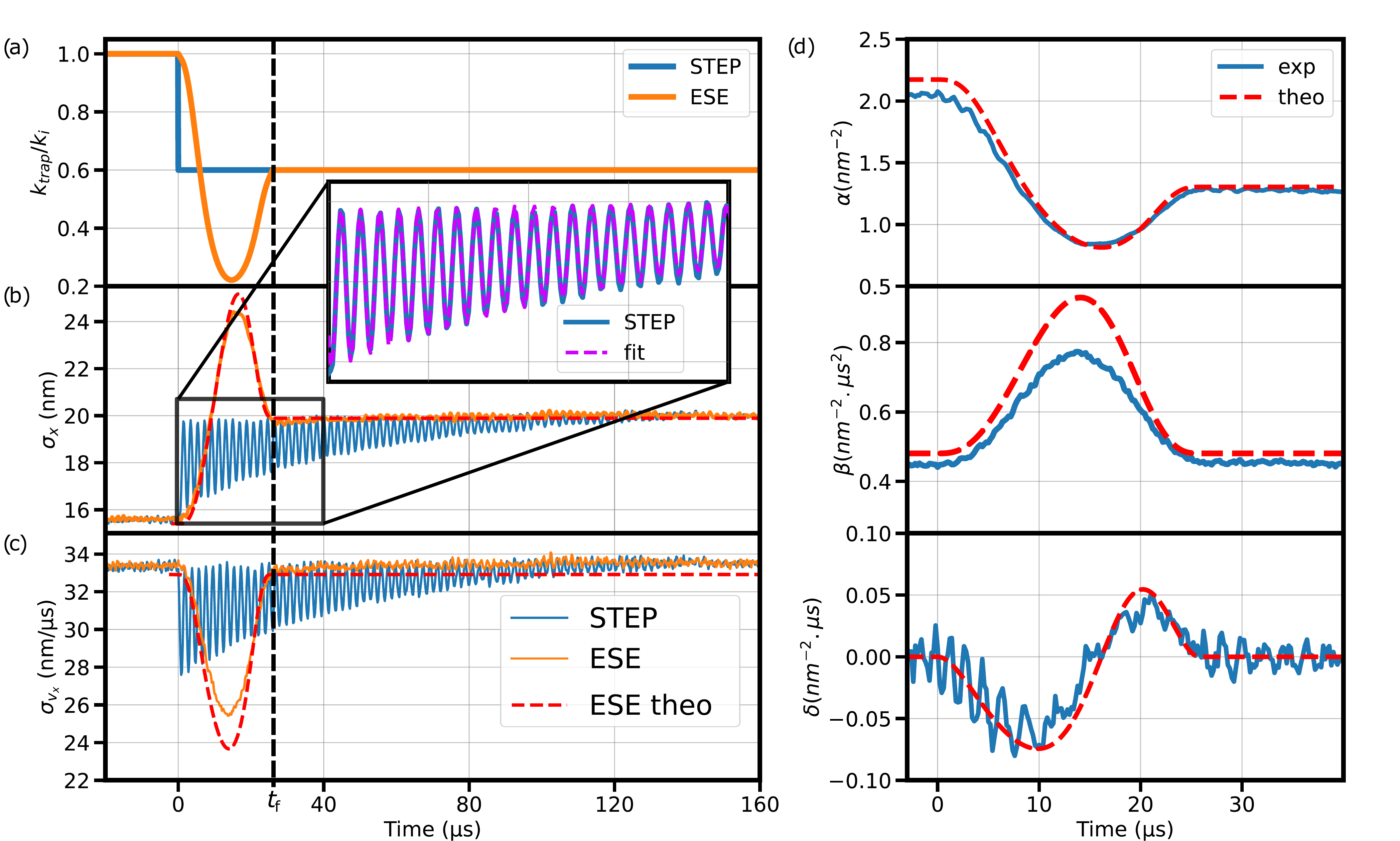}
\caption{ (a) Evolution of the trap stiffness for a harmonic expansion in the case of a STEP protocol (blue) and the accelerating ESE protocol described in the main text (orange).
(b) and (c) Evolution of the standard deviation in position $\sigma_{x}$ (b) and velocity $\sigma_{v_x}$ (c) for the STEP (blue) and the ESE protocol (orange) presented in (a). The black dotted vertical line corresponds to the final time $t_f=26$~µs of the ESE protocol. Dashed red lines are the expected theoretical values for $\sigma_{x}$ and $\sigma_{v_x}$
        Inset: zoom on  $\sigma_{x}$ highlighting the oscillations. A fit (dashed purple line) provides the value of $\omega_\text{relax}/2\pi =519 \pm 1$~kHz and $\Gamma/2\pi =3.1 \pm 0.2$~kHz.
(d) Evolution of $\alpha$, $\beta$ and $\delta$ (as defined in text) during the out-of-equilibrium regime of the ESE protocol pictured in (a). The experimental values (blue) are compared with those calculated for the ESE protocol (dashed red lines).}
\label{fig:comparison}
\end{figure*}

To determine the natural relaxation time of the system $t_\text{relax}$, we perform a STEP protocol, consisting of an instantaneous change in the trap stiffness $k_\text{trap}$ from an initial value $k_i$ to a final value $k_f$, as shown in Fig.~\ref{fig:comparison}-(a). We introduce the compression factor $\chi=k_f/k_i$. In the following, we focus on the case of isothermal expansion $\chi<1$, while our experimental setup can also naturally address isothermal compressions~\cite{SI}. 
We study the dynamics of the particle along the $x$-axis over a set of $2 \times 10^4$ isothermal expansions with $\chi=0.6$. Interestingly, the good signal-to-noise ratio of our measurement scheme allows us to determine the particle velocity $v_x$ from a point-by-point derivative of $x$. We thus study the particle relaxation after the STEP protocol by computing the standard deviation in position $\sigma_{x}$ and velocity $\sigma_{v_x}$,  as shown in Fig.~\ref{fig:comparison}-(b) and (c).
These data exhibit  a couple of interesting features. First, at equilibrium, we observe steady-state values  $\sigma_{x,\{i,f\}}^\text{eq} =\sqrt{k_B T / k_{\{i,f\}}}$ and $\sigma_{v_x}^\text{eq} = \sqrt{k_B T / m}$ as expected from the equipartition theorem. 
Besides, contrasting with the overdamped case~\cite{Martinez2016NP,Albay2020APL}, we observe in the transient regime damped oscillations in phase oppositions for $\sigma_x$ and $\sigma_{v_x}$. These oscillations can be seen as a coherent exchange of the system's average potential and kinetic energy, in analogy with a classical underdamped harmonic oscillator.  Finally, from a fit to the analytical solution, we confirm that in the strongly underdamped regime achieved here ($\Gamma\ll \omega_f$), the oscillation frequency is twice the natural trap frequency, $\omega_\text{relax} = 2 \omega_{x,f}=2\pi\times 519$~kHz  and that the system's characteristic relaxation time is given by the velocity relaxation time $t_\text{relax} = t_v = 1/\Gamma = 43$~µs, leading to a STEP equilibration in a time close to  $3t_\text{relax}$ (i.e. $\sigma_x$ and $\sigma_{v_x}$ reached 95\% of their equilibrium values)~\cite{SI}.
These observations starkly contrast with previous overdamped experiments, where the system thermalizes instantaneously with the environment, and in which the limiting timescale is given by the position relaxation time $t_x = \Gamma/\omega^2$ necessary for the nanoparticle to explore the new potential~\cite{Martinez2016NP, Albay2020APL}. They thus highlight the importance of addressing the underdamped regime to fully capture the nature of nanosystems.

Once this natural relaxation time is measured, we focus on shortcutting the system equilibration time. In that context, the theoretical proposal by Chupeau et al.~\cite{Chupeau_2018} is particularly interesting. Indeed, it extends the engineering of swift equilibration (ESE), initially developed for overdamped systems, to the underdamped regime.  
The idea behind this ESE formalism is to find a probability density function $\rho(x,v_x,t)$ that is a solution of the Fokker-Planck equation describing the dynamics of the particle,
\begin{equation}
        \dfrac{\partial \rho }{\partial t} +v_x \dfrac{\partial \rho }{\partial x} -\dfrac{k^x_\text{trap}}{m}x \dfrac{\partial \rho }{\partial v_x} = \dfrac{\Gamma}{m}\dfrac{\partial v_x \rho }{\partial v_x}+\dfrac{\Gamma k_B T}{m^2}  \dfrac{\partial^2 \rho }{\partial v_x^2}\, ,
        \label{eq:FP}
\end{equation}
and that reaches the final aimed equilibrium state in an arbitrary small finite time $t_f$. 
To this end, and by virtue of the linearity of the applied force, one may search for a Gaussian solution of the form
\begin{equation}
	\rho (x,v_x,t) = N(t)\exp(-(\alpha(t) x^2+\beta(t) v_x^2 + \delta(t) xv_x)) \, , 
\end{equation}
where $\alpha, \beta$ and $\delta$ are functions to be determined, and $N$ is a normalization factor~\cite{SI}. 
Solving this problem provides a continuous-time evolution of the trap stiffness $k^x_\text{trap}(t)$ as the control parameter of the system equilibration.
For example,  we depict in Fig.~\ref{fig:comparison}-(a), the trap stiffness evolution we compute for an ESE protocol corresponding to a five-fold acceleration over the nominal equilibration time $3t_\text{relax} = 129$~µs, and to a protocol duration $t_f=26$~µs.

Applying this ESE protocol to our particle leads to the evolution of the standard deviations $\sigma_x$ and $\sigma_{v_x}$  shown in orange in Fig.~\ref{fig:comparison}-(b) and (c). We verify that the system's relaxation is actually shortened and that both position and velocity variances reach their equilibrium values exactly at the protocol's final time $t_f=26$~µs (black dotted line in Fig.~\ref{fig:comparison}).
Furthermore, our ability to measure both $x$ and $v_x$ allows us to fully monitor the probability distribution function. Specifically, we can compute from these data the values of $\alpha$, $\beta$, and $\delta$ during the protocol. As shown in Fig.~\ref{fig:comparison}-(d), we observe a good agreement with the theoretical target functions enforced by the protocol, 
demonstrating the power of the ESE approach to engineer equilibration~\cite{SI}.
We also stress that the ESE strategy is a full feedforward approach. Thus, conversely to feedback, its efficiency does not rely on the  knowledge on the particle real-time dynamics.

These results are thus the first demonstration of accelerated equilibration of a particle in the underdamped regime. They stress that trap stiffness can be used as a single experimental control parameter  acting on both position and velocity to design shortcuts protocols. They also show that we can fully measure and reconstruct the evolution of the distribution of probability $\rho(x,v_x,t)$ during state-to-state transformations, highlighting levitated particles as a perfect system for out-of-equilibrium studies. 
Besides, within our approach, the set of admissible controls is infinite. Here we focus on shortcuts to equilibrium, but other protocols targeting minimum entropy production/dissipation~\cite{Muratore-Ginanneschi2014JSM,Muratore-Ginanneschi2014PRE} or mean work provided by the operator~\cite{Gomez-Marin2008JCP} could also be addressed. Characterizing such protocols requires retrieving the heat and work exchanged by the particle to the environment. Knowing the dynamics of the levitated particle, it is information that we can  retrieve~\cite{SI}.

Beyond accelerating equilibration, a natural question related to ESE protocols is their resilience to changes or uncertainties in the system's parameters.
Of particular interest is the design of robust protocols that shorten equilibration of a set of oscillators at different frequencies. Indeed, this could provide protocols insensitive to frequency fluctuations~\cite{Sansa2016} or able to address systems with multiple degrees of freedom, such as arrays of oscillators~\cite{Dago2020SP}, Brownian gyrators~\cite{Gyrator-baldassari,Frim2021PRE}, or multimode oscillators~\cite{LeCunuder2016}.

\begin{figure}[!htb]
\includegraphics[scale=0.5]{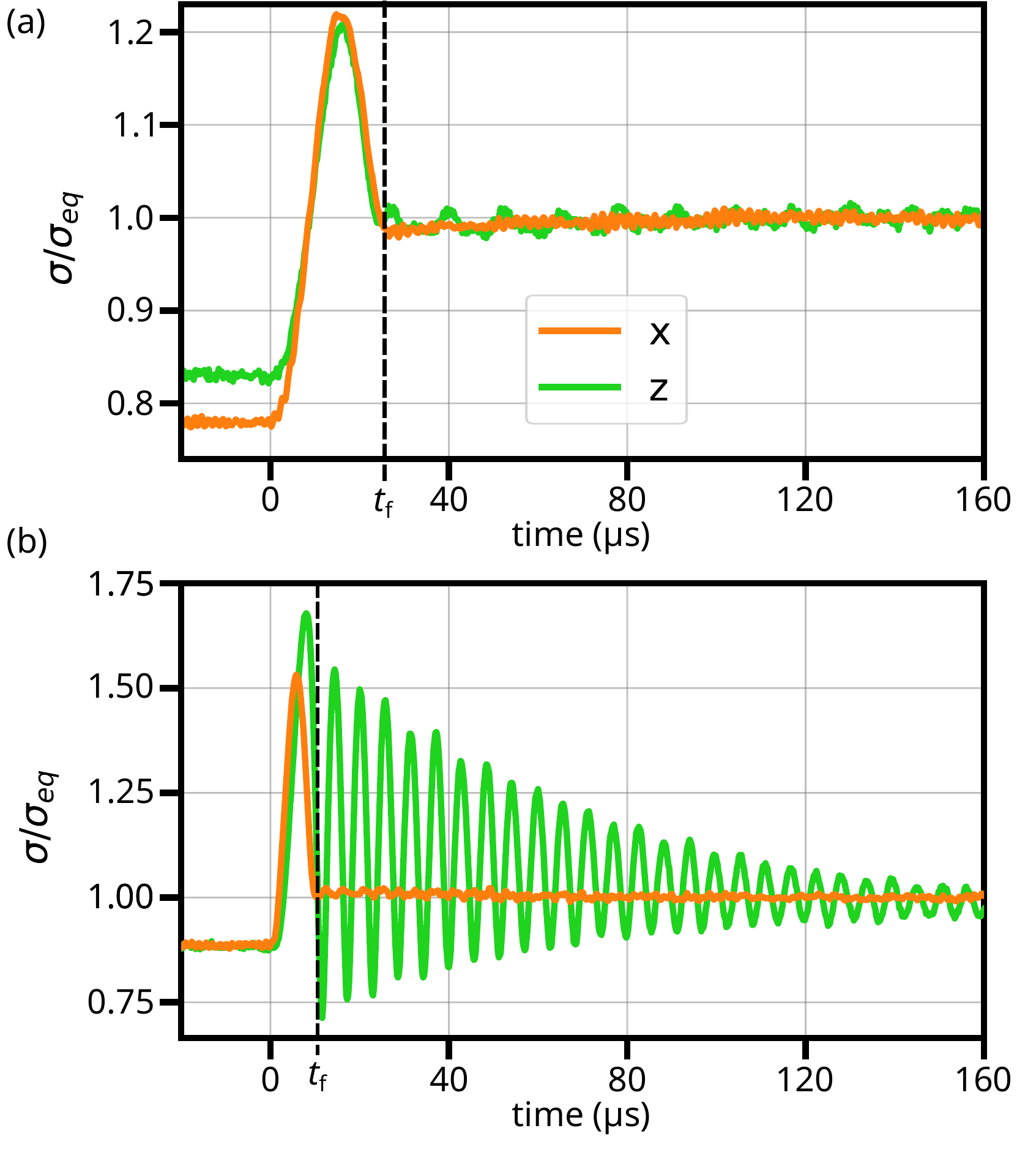}
\caption{Evolution of the position standard deviation $\sigma$, normalized by its final equilibrium value $\sigma_\text{eq}$, along the $x$-axis (orange) and $z$-axis (green) during an ESE protocol defined on the $x$-axis at the frequency $\omega_{x,i}/2\pi=340$~kHz, and with $\omega_z\approx \omega_x/3.4$. (a) Equilibration for a five-fold accelerating equilibration protocol for a $\chi=0.6$ expansion. (b) Equilibration for a 17-fold accelerating equilibration protocol for a $\chi=0.75$ expansion.
The black dotted vertical lines correspond to the final times $t_f=26$~µs (a) $t_f=7.75$~µs (b) of the two ESE protocols.}
\label{fig:x_vs_z}
\end{figure}

To test the robustness of the ESE protocols in our specific case, we take advantage of the three-dimensional nature of our system. Indeed, the three normal modes are non-degenerate and uncoupled to first order. Recording the three-dimensional dynamics of the particle allows us to study the system equilibration along different axes, corresponding to uncoupled oscillators at different frequencies undergoing the same relative stiffness evolution. 
Specifically, we focus on comparing $x$ and $z$-axes that represent the largest frequency mismatch in our experiment  $\omega_x/\omega_z{\approx 3.4}$. 
We thus apply the ESE protocols studied previously and designed for the $x$-axis (Fig.~\ref{fig:comparison}). To study the relaxation of each axis, we compute the standard deviation for the particle position along these $x$ and $z$ axes. As shown in Fig.~\ref{fig:x_vs_z}-(a), for this five-fold accelerated equilibration, the protocol designed for the $x$-axis works surprisingly well for the lower frequency $z$-axis, which also reaches equilibrium at the final protocol time $t_f=26$~µs. Note that similar results are also observed for the relaxation along the $y$-axis, that is at a natural frequency $\omega_y$ close from the one of the $x$-axis $\omega_x\approx1.3 \omega_y$~\cite{SI}. 

To push the limits of the robustness of the ESE protocols, we have also designed a faster protocol with a final time $t_f= 7.75$~µs, corresponding to a 17-fold accelerated equilibration compared to natural relaxation. 
The faster protocols are more demanding in terms of required total laser power. Due to experimental constraints, we thus reduce the expansion factor of the protocol to $\chi=0.75$.  
Fig.~\ref{fig:x_vs_z}-(b) shows the corresponding relaxation of the standard deviation of position along $x$ and $z$.
First, we verify that the protocol is efficient for the design axis ($x$-axis), and that $\sigma_x$ actually reaches its steady state value at the protocol's final time $t_f=7.75$~µs (black dotted line in the figure). We thus demonstrate that ESE protocols applied to a single oscillator can achieve equilibration accelerated by more than an order of magnitude in the underdamped regime.
Conversely, along the $z$-axis, the protocol drives the system out of equilibrium. After the protocol final time $t=t_f$, the system thus relaxes to equilibrium, dissipating its extra energy to the bath, as observed for STEP protocols in Fig.~\ref{fig:comparison}. We therefore observe oscillations of the position standard deviation $\sigma_z$ at the frequency $2\omega_{z,f}$ with a characteristic relaxation time $t_v=1/\Gamma$.

Hence, we observe that for moderate equilibration acceleration,  ESE protocols are resilient to changes in frequency. We can engineer equilibrium states even far from the target frequency. 
When the protocol speed is increased, this property becomes harder to meet. For the fastest protocols studied here,  to observe an efficient equilibration, the actual system frequency must match the one of the protocol within a few percents~\cite{SI}.

To conclude, we have shown that engineering swift equilibration can be successfully applied in the underdamped regime. We demonstrate accelerated equilibration by more than an order of magnitude for expansion protocols for a single frequency oscillator, and intermediate speed-up for a multidimensional one. In addition, we highlight that by using optically levitated particles, we can reconstruct the particle's dynamics, providing complete information about the system's probability density function. Taking advantage of the levitated particle's different degrees of freedom,  we also address our shortcuts to equilibrium protocols' robustness. Finally, we report that moderate acceleration can be applied to oscillators with a broad frequency range. 
Our work paves the way for developing generic state-to-state protocols and discussing the fundamental physical limits of shortcuts to equilibrium protocols' robustness. Thus, future works may include fast multidimensional shortcuts, taking advantage of the vast space of relaxation paths offered by ESE-like protocols~\cite{Chupeau_2018,Li2017PRE,Chupeau2018PRE,Frim2021PRE} or by optimal control theory framework~\cite{Martikyan2020}. 
Also, by taking advantage of the recent developments of thermal bath engineering~\cite{Militaru2021NC,Rademacher2022PRL} and fine temperature control~\cite{Magrini2021N,Tebbenjohanns2021N} of levitated particles, a natural extension rests in the use of shortcuts to equilibrium and optimal protocols for the optimisation of power and efficiency of nano-heat engines~\cite{Dechant2015PRL}.

\begin{acknowledgments}
        This work is supported by the Investissements d’Avenir of LabEx PALM (ANR-10-LABX-0039-PALM), and by the ANR  projects OPLA (ANR-20-CE30-0014) and STATE (ANR-18-CE30-0013).
        We thank C. Lopez and S. L'Horset for support with electronics, F. Bretenaker for lending a laser, H. Huet and B. Gebel for seminal experimental work, and M. Baldovin, L. Bellon, S. Ciliberto, S. Dago, C. Plata together with A. Prados for useful discussions.
\end{acknowledgments}

\bibliography{ESE}

\end{document}



\title{Supplementary Information for Shortcuts to equilibrium with a levitated particle in the underdamped regime}

\author{Damien Raynal}
\author{Timothée de Guillebon}%
\affiliation{Université Paris-Saclay, ENS Paris-Saclay, CNRS, CentraleSupélec, LuMIn, 91405 Orsay Cedex, France}
\author{David Guéry-Odelin}
\affiliation{Université Paul Sabatier - Toulouse 3, CNRS, LCAR, 31062 Toulouse Cedex 9, France}
\author{Emmanuel Trizac}
\affiliation{Université Paris-Saclay, CNRS, LPTMS, 91405 Orsay Cedex, France}
\author{Jean-Sébastien Lauret}%
\author{Loïc Rondin}%
\email{loic.rondin@universite-paris-saclay.fr}
\affiliation{Université Paris-Saclay, ENS Paris-Saclay, CNRS, CentraleSupélec, LuMIn, 91405 Orsay Cedex, France}

\date{\today}

\maketitle


\section{Experimental Methods}

The particle is optically trapped at the focus of a NA=0.85 objective (Olympus LCPLN100XIR) using a high-power near-infrared laser beam (Coherent Mephisto MOPA up to 7~W, $\lambda=1064$~nm). 
The laser power is tuned using an acousto-optics Modulator (AA Opto-Electronic MT110-A1.5-IR) driven with a fast arbitrary wave generator (Spectrum Instrumentation M4i.6621-x8).
The particle dynamics are measured with a common path interferometer using an ancillary laser beam (Laser Quantum GEM, $\lambda_\text{meas}=532$~nm, $P_\text{meas}\approx 7$~mW) and a quadrant photodetector (Hamamatsu S4349). The 3D dynamics of the particle is recorded onto a digital scope (Picoscope 4824A)  at 5~MSamples/s. The particle position $x(t)$ is corrected by subtracting its mean value at equilibrium after each protocol realization, to eliminate experimental drifts.

\section{Damping $\Gamma$ of levitated particles}
\label{sec:damping}

For a spherical particle of radius $r$ in a rarefied gas, the damping rate $\Gamma$ is directly proportional to the gas pressure  $p_\text{gas}$~\cite{Gieseler2012}:
\begin{equation}
\Gamma = 0.619 \dfrac{9}{\sqrt{2\pi}\rho_\text{SiO$_2$}}\sqrt{\dfrac{M}{N_A k_B T}}\dfrac{p_\text{gas}}{r}\, ,
\end{equation}
where $\rho_\text{SiO$_2$}\approx 2200$~kg/m$^3$ is the silica density, $M$ the molar mass of air, $T$ the environment temperature.

Considering the silica particle used in the main text, of expected radius $r=73$~nm as given by the provider (microParticles GmbH), we find a damping rate $\Gamma_\text{theo}=2\pi\times3.8 $~kHz at a pressure $p_\text{gas}=5$~hPa.

Experimentally, this damping can be measured for pressures above a few hPa directly from the linewidth of the particle dynamics power spectral density~\cite{Gieseler2012}, or alternatively, as discussed in the main text, from the relaxation time in a STEP protocol.

Using the second approach, we find  $\Gamma_\text{exp}=2\pi \times 3.1$~kHz.
This result is in good agreement with the expected value $\Gamma_\text{theo}$. 

\newpage 
\section{Overdamped ESE protocols}
\begin{figure}[htbp]
        \centering
        \includegraphics[width=.5 \textwidth]{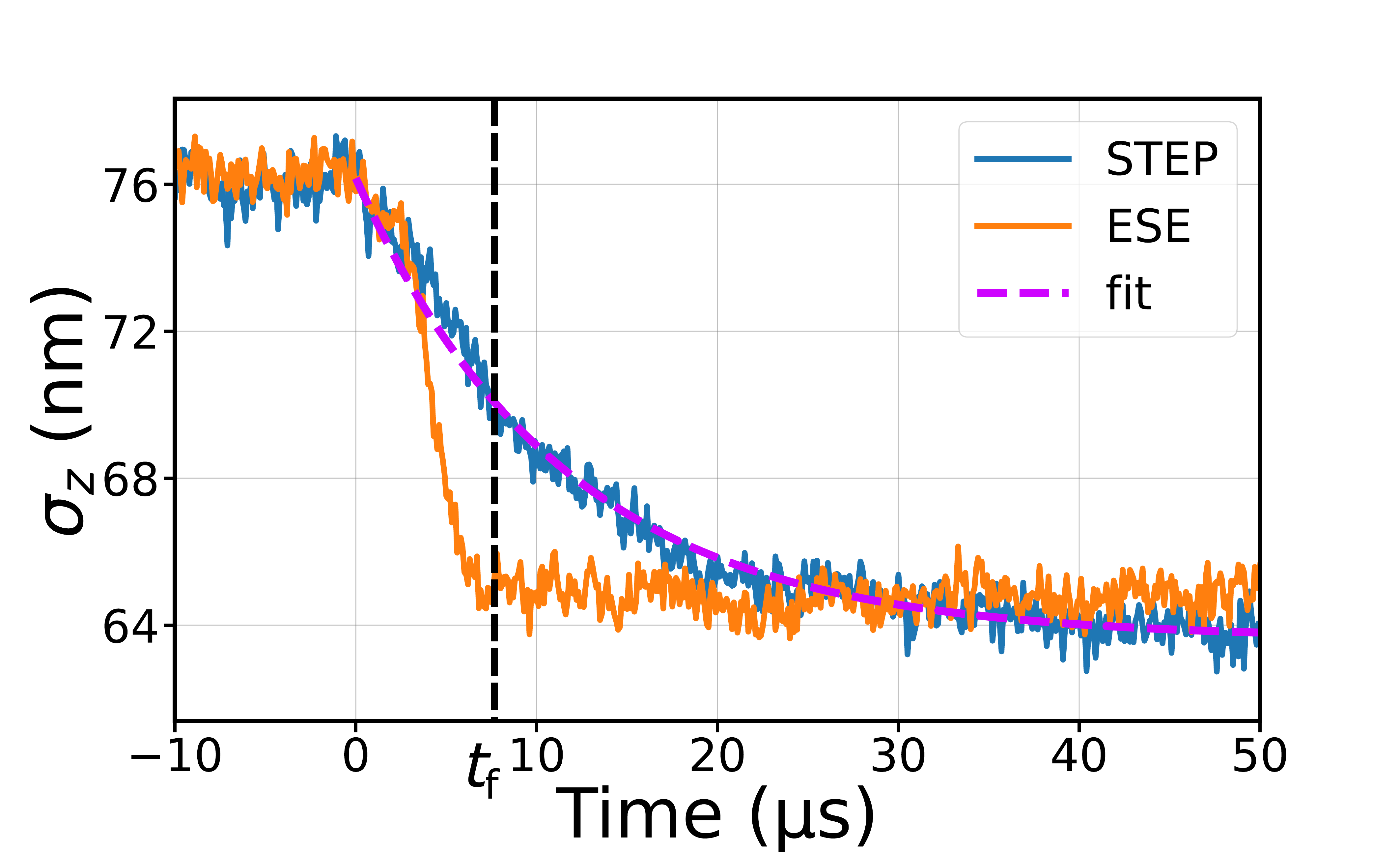}
        \caption{Overdamped STEP (blue) and ESE (orange) compression protocols for the $z$-axis of an optically trapped particle in air.  The purple dashed line correspond to an exponential fit, giving a relaxation time $t_z=11.5$~µs.}
        \label{fig:overdampedESE}
\end{figure}
As mentioned in the main text, and detailed in the Supplementary Note S1,
our experimental apparatus allows to easily change the damping condition by tuning the gas pressure inside the vacuum chamber. 
This allows to address ESE protocols for a broad range of damping conditions. Specifically,  it is possible to study overdamped protocols~\cite{Martinez2016NP}. As a proof of principle, figure~\ref{fig:overdampedESE} shows the case of an overdamped compression with $\chi=1.4$ for the $z$-axis, realized at ambient pressure $p_\text{gas}=10^5$~hPa ($\Gamma=2\pi\times 570$~kHz $\gg \omega_z^f=2\pi\times 92$~kHz). The reference STEP protocol (in blue) shows the expected position relaxation time toward equilibrium $t_z=\frac{\Gamma}{(\omega_z^f)^2}$. The ESE protocols (orange) allows us to shortcut the equilibration time by a factor 4, to $t_f=8$~µs.

\newpage 
\section{ESE protocols for compression}
\begin{figure}[htbp]
        \centering
        \includegraphics[width=.95 \textwidth]{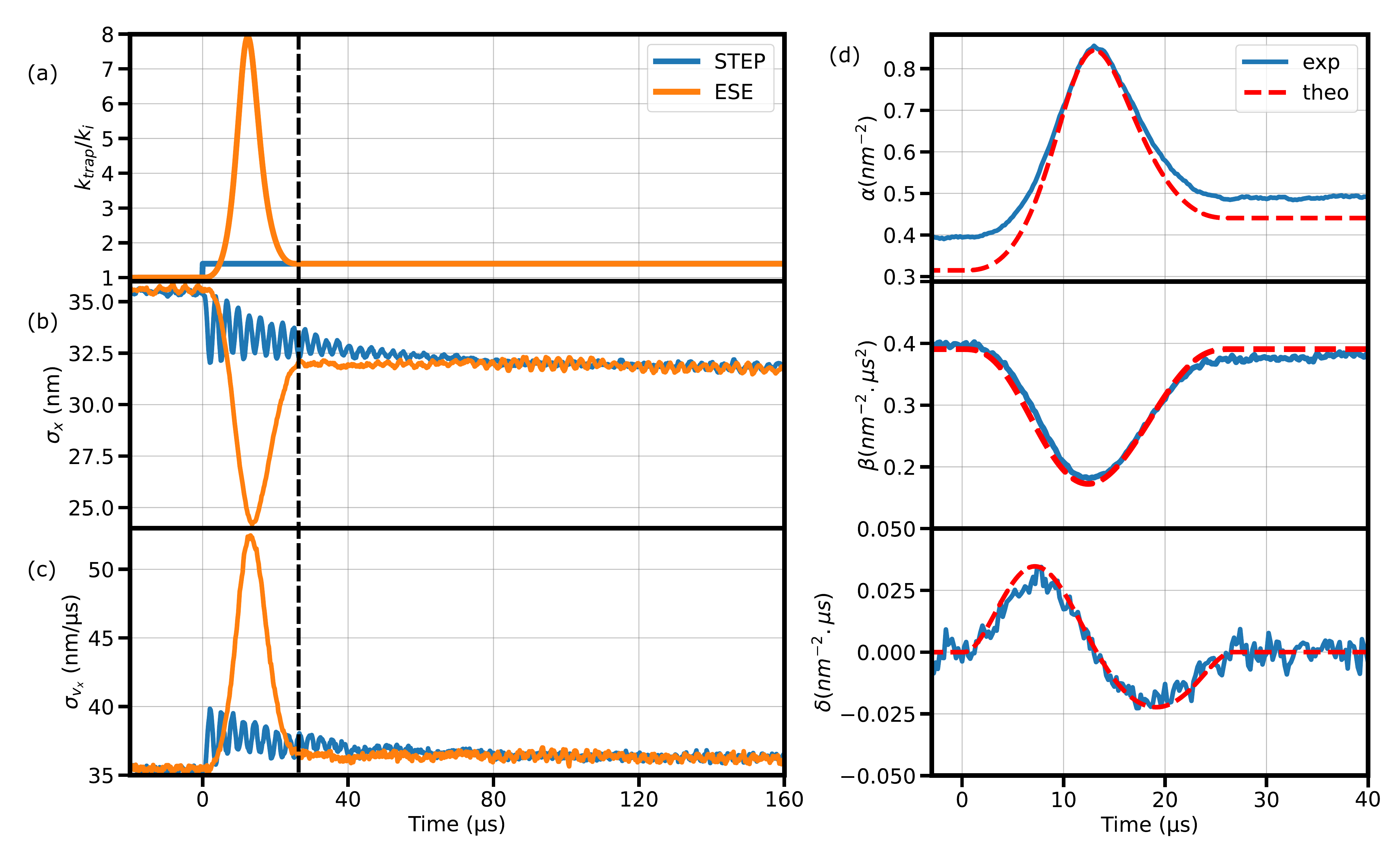}
        \caption{ (a) Evolution of the trap stiffness for a harmonic compression in the case of a STEP protocol (blue line) and an ESE protocol corresponding to $\chi=1.4$ and $t_f=26$~µs (5-fold speed-up)(orange).
        (b) and (c) Evolution of the standard deviation in position $\sigma_{x}$ (b) and velocity $\sigma_{v_x}$ (c) for the STEP (blue) and the ESE protocol (orange) presented in (a). The black dotted vertical line corresponds to the final time $t_f=26$~µs of the ESE protocol.
        (d) Evolution of $\alpha$, $\beta$ and $\delta$ (as defined in text) during the out-of-equilibrium regime of the ESE protocol pictured in (a). The experimental values (blue) are compared with those calculated for the ESE protocol (dashed red line).}
        \label{fig:Compression}
\end{figure}

\newpage 
\section{Evolution of variance for STEP protocols}
\noindent Let's consider a 1D harmonic potential, along the $x$ axis, defined by its angular frequency $\omega(t)$. A particle of mass $m$ is trapped in the harmonic potential at a temperature $T_0$, and the system damping is $\Gamma$. We introduce 
\begin{equation*}
\left\{  
        \begin{aligned}
                \sigma_{xx} &=& \langle x^2\rangle -\langle x\rangle ^2 \\
                \sigma_{xv} &=& \langle xv\rangle -\langle x\rangle \langle v\rangle \\
                \sigma_{vv} &=& \langle v^2\rangle -\langle v\rangle ^2
        \end{aligned}
\right.
\end{equation*}
These quantities are coupled through the linear system~\cite{Zerbe1994PRE}:
\begin{equation}
        \dfrac{\mathrm d}{\mathrm d t} \begin{pmatrix}
                \sigma_{xx}\\
                \sigma_{xv}\\
                \sigma_{vv}
\end{pmatrix} = 
\begin{pmatrix}
        0 & 2 & 0\\
        -\omega^2 & -\Gamma & 1 \\
        0 & -2\omega^2  & -2\Gamma
\end{pmatrix}
 \begin{pmatrix}
                \sigma_{xx}\\
                \sigma_{xv}\\
                \sigma_{vv}
\end{pmatrix}
+  \begin{pmatrix}
                0\\
                0\\
                \dfrac{2k_B T_0 \Gamma}{m}
\end{pmatrix} 
\end{equation}

We address a STEP protocol, where the trap frequency is suddenly changed from $\omega_i$ to $\omega_f=\sqrt{\chi} \omega_i$ at $t=0$. Following the main text notation, $\chi$ is the expansion factor. 
Solving this set of equations for the initial conditions $\sigma_{xx}(0)=\sigma_i$ and $\sigma_{vv}(0)=\sigma_i \omega_i^2$ leads to
\begin{equation}
                \sigma_{xx}(t)=\sigma_i \dfrac{\chi -1}{\chi}\left[ \dfrac{2 \omega_f^2}{\tilde \Omega^2}+\dfrac{2 \omega_f^2-\Gamma^2}{\tilde \Omega^2}\cos \tilde \Omega (t-t_0) + \dfrac{\Gamma}{\tilde \Omega} \sin \tilde \Omega (t-t_0) \right] e^{-\Gamma (t-t_0)} + \dfrac{\sigma_i}{\chi}\, ,
        \label{eq:GenSol}
\end{equation}
with $\tilde \Omega =\sqrt{4\omega_f^2-\Gamma^2}$.
We note, that for the deep underdamped regime $\omega_f\gg \Gamma$, as verified in the main text, then 
\begin{equation}
                \sigma_{xx}(t)=\sigma_i \dfrac{\chi -1}{2\chi}\left[ 1+\cos\left( 2\omega_f t\right) \right] e^{-\Gamma t} + \dfrac{\sigma_i}{\chi} \, . 
        \label{eq:GenSolApprox}
\end{equation}
The oscillation frequency observed during the STEP protocols is then twice the natural final frequency of the trap $\omega_f$.
\begin{figure}[htbp!]
        \centering
        \includegraphics[width=0.8\textwidth]{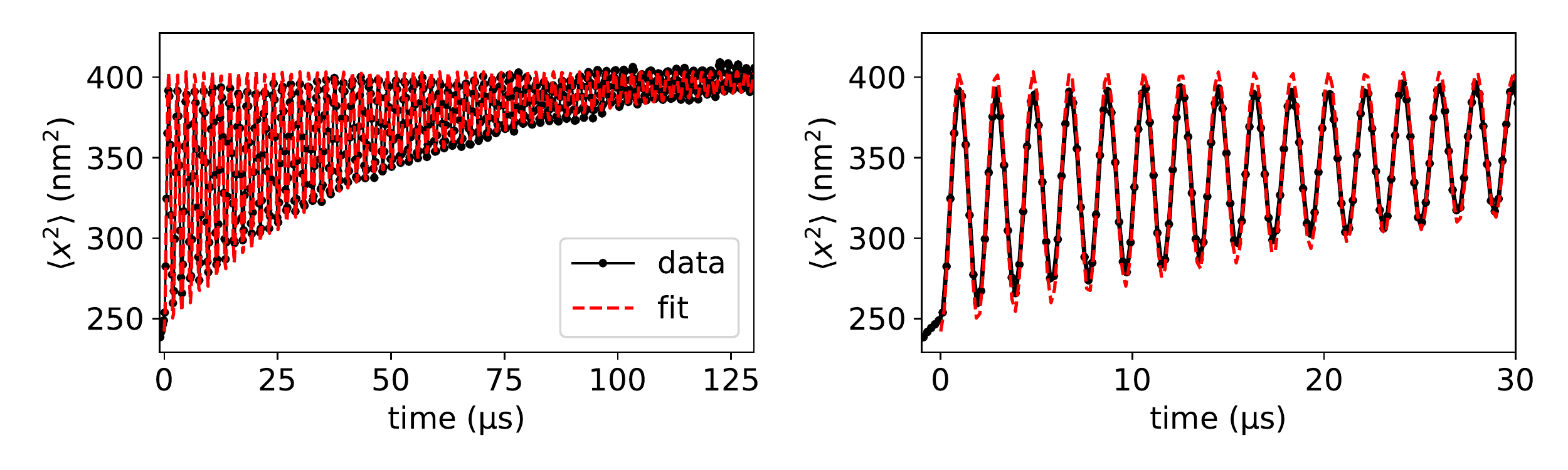}
        \caption{Position variance $\sigma_{xx}$ (black line) during a STEP protocol and associated fit. The fit allows us to determine the values of $\omega_\text{relax}=2\omega_f$ and $\Gamma$. The timebase has been corrected to enforce $t_0=0$~µs. }
        \label{fig:fit_sigxx}
\end{figure}

Experimentally, we fit the experimental data computed for the position variance $\sigma_{xx}=\sigma_x^2$ using equation~(\ref{eq:GenSol}), with $\Gamma$ and $\omega_f$ as free parameters. The results are presented in figure~\ref{fig:fit_sigxx}. In the main text (Figure 2(b)), we plot the square-root counterpart, to depict the position standard deviation $\sigma_x$.

\newpage 
\section{Kurtosis for STEP and ESE protocols}
    \begin{figure}[htp!]
        \centering
        \includegraphics[width=0.8\textwidth]{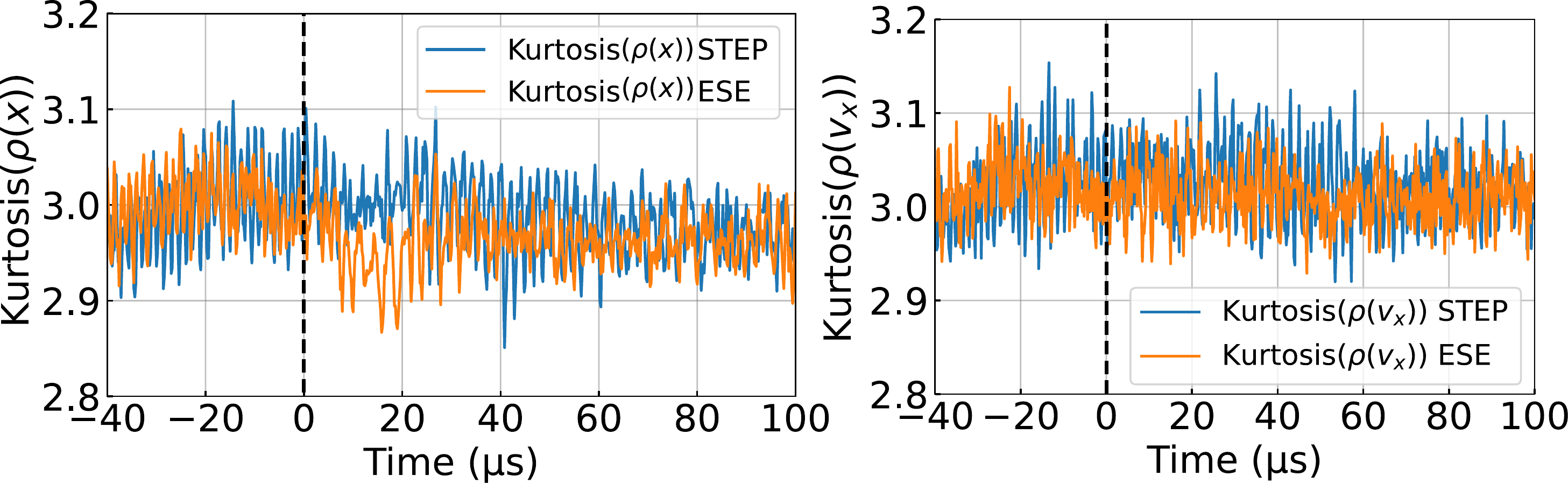}
        \caption{Kurtosis of $\rho(x)$ and $\rho(v_x)$ during STEP and ESE protocols}
        \label{fig:Kurtosis}
    \end{figure}
    The kurtosis of the particle distribution along the position $x$ axis $\rho(x)$ and velocity axis $\rho(v_x)$ during STEP and ESE protocols is shown in figure~\ref{fig:Kurtosis}. The measured kurtosis remains very close from 3, demonstrating that, if existing, deviation from Gaussianity remains small.

\section{Link between the ESE variables and the velocity and position variance}

The shortcut protocols used in the main text enforce a probability density for the system of the form 
\begin{equation}
        \rho(x,v,t) = N_0 \exp(-(\alpha(t)x^2 + \beta(t)v^2 + \delta(t)xv))
\end{equation}
For such a distribution, one can compute the quantities $\sigma_{xx}$, $\sigma_{vv}$ and $\sigma_{vx}$. We can thus determine the value of $\alpha,\beta$ and $\delta$~:
\begin{equation*}
\alpha =\frac{\sigma_{vv}}{2( \sigma_{vv}\sigma_{xx}-\sigma_{xv}^2)},
\quad
\beta =\frac{\sigma_{xx}}{2( \sigma_{vv}\sigma_{xx}-\sigma_{xv}^2)}
\end{equation*}
\begin{equation*}
\delta =\frac{-\sigma_{xv}}{( \sigma_{vv}\sigma_{xx}-\sigma_{xv}^2)}
\end{equation*}

In the main text (Figure 2(d)), we thus reconstruct the PDF of the system during the ESE process, by computing the values of   $\alpha,\beta$ and $\delta$ from the position variance $\sigma_{xx}$, the velocity variance $\sigma_{vv}$ and the cross-correlated term $\sigma_{xv}$.

Conversely, if $\alpha, \beta$ and $\delta$ are known, one can compute the values of $\sigma_x$ and $\sigma_{v_x}$ as shown in figures 2b and c of the main text.

\section{Definition of the ESE protocols}

\subsection{The ``protocol A'' solution}

The ESE protocols used in the main text are the type defined as "\textit{protocol A}" in reference~\cite{Chupeau_2018}. The idea behind this work is to engineer the evolution of the distribution 
\begin{equation}
	\rho (x,v_x,t) = N(t)\exp(-(\alpha(t) x^2+\beta(t) v_x^2 + \delta(t) xv_x)) \, , 
\end{equation}
via the control parameter $k_{trap}$.
Injecting this Ansatz into the Fokker-Planck equation:
\begin{equation}
        \dfrac{\partial \rho }{\partial t} +v_x \dfrac{\partial \rho }{\partial x} -\dfrac{k_\text{trap}}{m}x \dfrac{\partial \rho }{\partial v_x} = \dfrac{\Gamma}{m}\dfrac{\partial v_x \rho }{\partial v_x}+\dfrac{\Gamma k_B T}{m^2}  \dfrac{\partial^2 \rho }{\partial v_x^2}\, ,
        \label{eq:FP}
\end{equation}
provides a set of non-linear equations linking the rescaled functions 
$$
 \left\{
    \begin{array}{lcl}
            \tilde{\kappa} &=& \dfrac{k_\text{trap}}{k_i} \\
            \tilde{\alpha} &=& \dfrac{2k_B T}{k_i}\alpha \\
            \tilde{\beta} &=& \dfrac{2k_B T}{k_i}\beta \\
            \tilde{\delta} &=& \dfrac{2k_B T}{k_i}\delta
    \end{array}
\right.
$$
expressed  in rescaled time $s=t/t_f$.

Introducing the auxiliary quantity 
\begin{equation}
        \tilde{\Delta} = (\tilde{\alpha}-\frac{\tilde{\delta}^2}{\tilde{\beta}})\tilde{\beta} \, ,
        \label{eq:Delta_def}
\end{equation}
allows rewriting the aforementioned functions in terms of $\tilde \Delta$. 
Thus, to define a protocol, all that is needed is to find a $\tilde{\Delta}$ which satisfies the boundary conditions on $\tilde{\alpha}$, $\tilde{\beta}$ and $\tilde{\delta}$, in particular $\tilde{\alpha}(1) = \tilde{\Delta}(1) = \chi$. In reference~\cite{Chupeau_2018}, to ensure continuity of the control parameter $k_\text{trap}$, the two first derivative of $\tilde \Delta$ are taken to be $0$ in $s=0$ and $s=1$.

Finally, the control parameter $k_\text{trap} = k_i \cdot \tilde{\kappa}$ is then fully defined as:
\begin{equation}
        \tilde{\kappa} = \frac{\dot{\tilde{\alpha}}}{2 \omega_i \tilde{\delta}} + \frac{\Gamma}{\omega_i}\tilde{\delta} \, ,
        \label{eq:kappa}
\end{equation}
Following Chupeau et al.~\cite{Chupeau_2018}, we look for a polynomial solution for $\tilde{\Delta}$. The lower order admissible polynomial is then: 
\begin{equation}
\tilde{\Delta}(s) = 1+(\chi-1)(35s^4-84s^5+70s^6-20s^7) \, .
\label{eq:Delta_pol}
\end{equation}
All the ESE protocols shown in the main text are based on this approach.

\subsection{Alternative ESE protocols}
Nevertheless, we note that the choice of $\tilde{\Delta}(s) $ is arbitrary, and one could imagine using a different function, a higher order polynomial, or a different basis for the decomposition. 
For instance, one can use a sinusoidal basis. The lower admissible order then leads to 
\begin{equation}
        \tilde{\Delta}_\text{sin}(s) = \frac{1+\chi}{2}+9\frac{1-\chi}{16}\cos(\pi s) -\frac{1-\chi}{16} \cos(3\pi s) \, .
\label{eq:Delta_sin}
\end{equation}
Experimentally, using this protocol provides similar results to those obtained with the polynomial \emph{Protocol A} described previously.

\newpage 
\section{Heat and Work for the used protocols}

\begin{figure}[h]
        \centering
        \includegraphics[width=\textwidth]{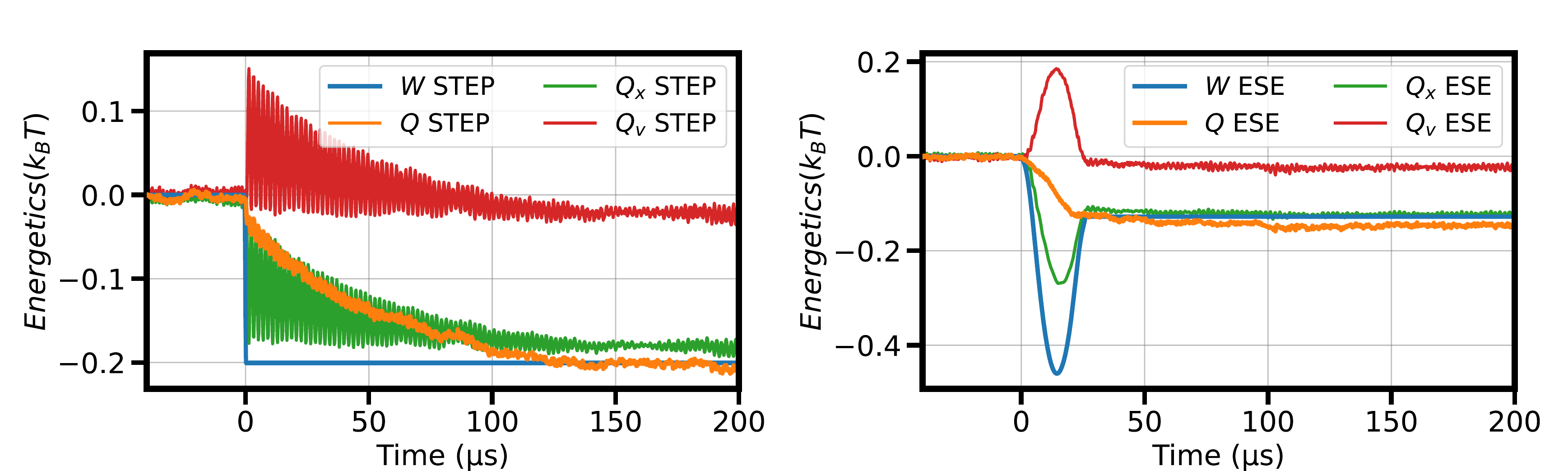}
        \caption{Heat and work for a STEP protocol (a) and the equivalent shortcut ESE protocol presented in the main text (b). The work for the STEP protocol is obtained from Eq.~\eqref{eq:meanWork}.}
        \label{fig:WandQ}
\end{figure}

From the particle time traces, we can compute the cumulative heat 
\begin{equation}
\langle Q (t) \rangle =  -\int_0^t k(t') \langle x  v_x\rangle \mathrm d t' - \left[\frac{1}{2}m \langle v_x^2 \rangle \right]_0^t = \langle Q_x (t) \rangle + \langle Q_v (t) \rangle \, ,
\label{eq:Q}
\end{equation}
and the cumulative work 
\begin{equation}
\langle W(t) \rangle = \int_0^t \dfrac{1}{2}\dot k(t') \langle  x^2 \rangle \mathrm d t' \, ,
\label{eq:W}
\end{equation} exchanged between the system and the environment for the STEP and the shortcut protocol presented in figure 2 of the main text.  The results are shown in figure~\ref{fig:WandQ}. 

In the case of a STEP protocol, the cumulative work is estimated from its theoretical value 
\begin{equation}
\langle W(t) \rangle=\dfrac{\chi -1}{2}k_B T 
\label{eq:meanWork}
\end{equation}
for any positive time. 

\newpage

\section{Relaxation along the $y$-axis}
\begin{figure}[htbp]
        \centering
        \includegraphics[width=.5 \textwidth]{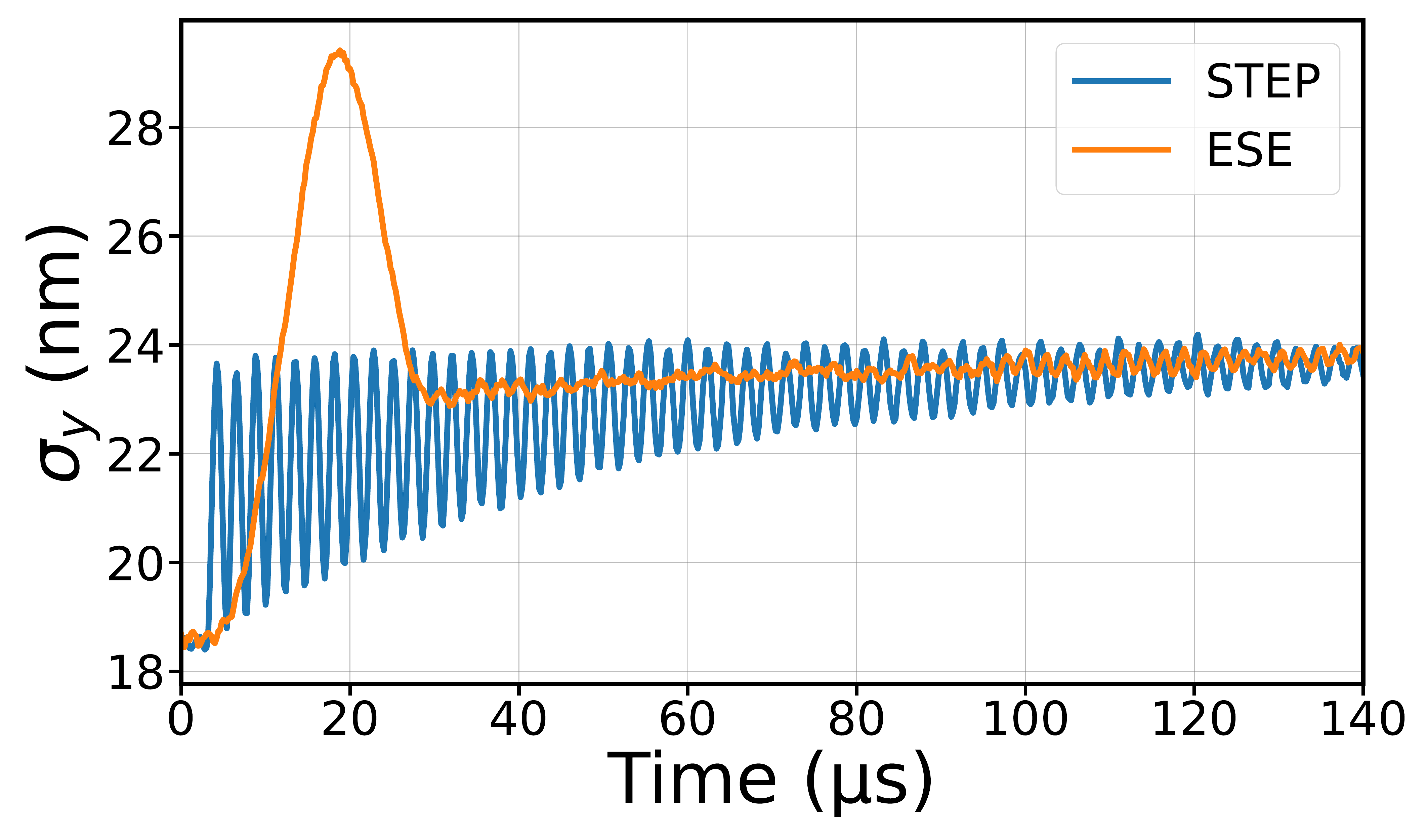}
        \caption{Relaxation along the $y$-axis for the STEP (blue) and ESE (orange) protocols, the latter targeting acceleration equilibration along the $x$-axis to $t_f=26$~µs, as shown in figure~3a in the main text. As for the $z$-axis, relaxation to equilibrium is observed in the target time for moderate speed-up (here 5-fold).}
        \label{fig:yaxis}
\end{figure}

\section{State-to-state protocols robustness}

To address the robustness of a transformation protocol defined for a reference system, we propose to characterize how close from equilibrium ends an arbitrary system submitted to this protocol. 

To characterize this distance to equilibrium between the system distribution at the end of the protocol  $p(x,v)$ and the equilibrium distribution $q(x,v)$, we use the Kullback–Leibler divergence $D(p||q)$. This estimator is defined as
\begin{eqnarray}
        D(p||q) &=& \displaystyle \iint \mathrm d x \mathrm d v\,  p(x,v) \ln\left(\dfrac{p(x,v)}{q(x,v)}\right)\, ,
\end{eqnarray}
and can be seen as a statistical distance between the two distributions $p$ and $q$.

In the present paper, we are interested in the efficiency of an ESE protocol defined for a system of natural frequency $\omega_\text{ref}$ applied to a system of frequency $\omega$ (the second axis in the main text). We thus determine the ESE protocol corresponding to a system of natural frequency $\omega_\text{ref}$ and damping $\Gamma$, and for a final time $t_f$. 
We then numerically compute the evolution of the distribution 
\begin{equation}
\rho(x,v_x,t) = \dfrac{\sqrt{4 \alpha(t)\beta(t)-\delta(t)^2}}{2\pi}e^{-\alpha(t) x^2 -\beta(t) v^2 - \delta(t) xv}
\end{equation}
of a system of frequency $\omega$ submitted to the protocol (same damping, same final time). 
This is done by numerical integration of the set of equations (14) in reference~\cite{Chupeau_2018}.
We thus compute the final value of the distribution $p(x,v) = \rho(x,v_x,t_f)$

By considering the equilibrium target distribution
\begin{equation}
   q(x,v)=\rho_\text{eq}(x,v_x,t_f) = \dfrac{m\omega \chi}{\pi k_B T}  \exp{\left(-\chi \frac{m\Omega_i^2}{k_b T} x^2 - \frac{m}{k_b T}v^2\right)} \, , 
\end{equation}
one can write the Kullback-Leibler divergence as:
\begin{eqnarray}
        D(p||q) 
                &=& \dfrac{1}{2}\ln \left[\dfrac{4\alpha\beta-\delta^2}{4\alpha^\text{eq}\beta^\text{eq}}\right] - (\alpha-\alpha^\text{eq})\sigma_{xx}^f - (\beta-\beta^\text{eq})\sigma_{vv}^f - \delta \sigma_{xv}^f\, ,
\end{eqnarray}
where 
$$
\left\{
\begin{array}{llllr}
        \sigma_{xx} &=& \langle x^2 \rangle &=& \dfrac{2 \beta}{4\alpha\beta -\delta^2} \\
        \sigma_{vv} &=& \langle v^2 \rangle &=& \dfrac{2 \alpha}{4\alpha\beta -\delta^2} \\
        \sigma_{xv} &=& \langle xv \rangle &=& \dfrac{-\delta}{4\alpha\beta -\delta^2} \\
\end{array}
\right.
$$

\begin{figure}[htbp]
        \centering
        \includegraphics[width=0.8\textwidth]{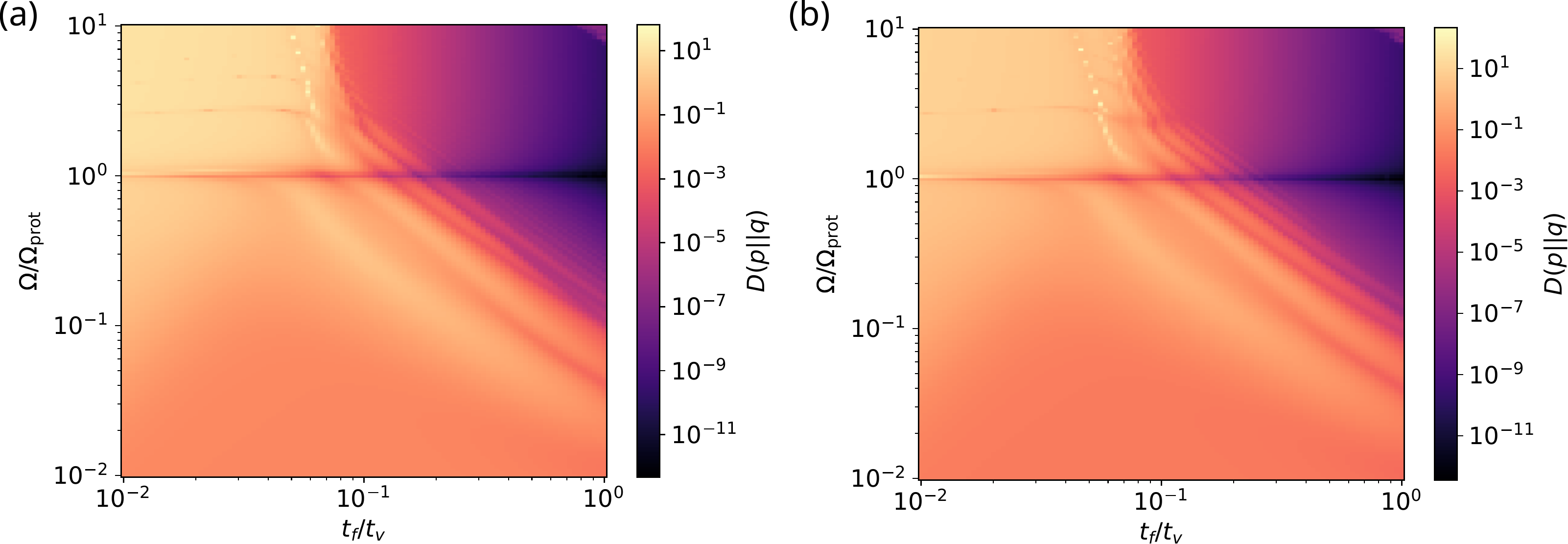}
        \caption{Kullback-Leibler divergence for the ESE protocols as a function of the final time $t_f$ and the system frequency  $\omega$. (a) Protocol used in the main text. (b) Alternative sinusoidal protocol defined by equation~\ref{eq:Delta_sin}.}
        \label{fig:Kullback}
\end{figure}

The Kullback-Leibler divergence for the ESE protocols as a function of the final time $t_f$ and the frequency difference $\omega/\omega_\text{ref}$ is shown in figure~\ref{fig:Kullback}-(a). 

A couple of interesting features are observed. First, the protocols are more robust for moderate acceleration, as discussed for experimental data in the main text. Then, the evolution of the  Kullback-Leibler divergence is non-monotonic, and we observe a coupling between frequency and final time impacting protocol robustness. As a consequence, for a given system frequency $\omega$, tuning the final time to reach a state closer to equilibrium could be interesting. 

To discuss the universality of these features, we apply the same procedure for the protocols defined in the previous part, which use a sinusoidal decomposition for the $\Delta$ function (see equation~\ref{eq:Delta_sin}). 
The results are presented in figure~\ref{fig:Kullback}-(b), demonstrating the same properties, with a worse protocols efficiency for decreasing final time and the non-monotonic behaviour of the Kullback-Liebler divergence.

Finally, we demonstrate that Kullback-Leibler divergence could be a strategy to characterize the robustness of state-to-state protocols. 
If the two proposed protocols share the same limitation to moderate speed-up, our strategy could be used to discuss other protocols, both for swift equilibration and optimization protocols~\cite{Muratore-Ginanneschi2014PRE}.




\bibliography{ESE}